\documentclass[prl,twocolumn,showpacs,preprintnumbers,amsmath,amssymb]{revtex4}
\usepackage{graphicx}
\usepackage{dcolumn}
\usepackage{bm}

\newcommand{\bnll}[1]{\begin{mathletters}\label{#1}\begin{eqnarray}}
\newcommand{\enll}{\end{eqnarray}\end{mathletters}}

\begin{document}







\title{Nuclear Tetrahedral Symmetry: 
       Possibly Present Throughout the Periodic Table\thanks{This 
       work has been partly supported by the Program of Scientific
       Exchange between the IN$_2$P$_3$, France, and Polish Nuclear Research
       Institutions, and by a grant from POLONIUM.}
       }
       
\author{ J.~Dudek$^1$, A.~Go\'zd\'z$^{1,2}$, N.~Schunck$^1$ 
         and M. Mi\'skiewicz$^{2}$ }

\affiliation{
             {\it $^1$Institut de Recherches Subatomiques 
                  IN$_2$P$_3$-CNRS/Universit\'e Louis Pasteur \\
                  F-67037 Strasbourg Cedex 2, France}\\
             {\it $^2$Katedra Fizyki Teoretycznej  
                  Uniwersytet Marii Curie-Sk\l{}odowskiej,}\\
             {\it PL-20031 Lublin, Poland}
        }


\begin{abstract}

More than half a century after the fundamental, spherical shell structure in 
nuclei has been established, theoretical predictions indicate that the 
shell-gaps comparable or even stronger than those at spherical shapes may exist. 
Group-theoretical analysis supported by realistic mean-field calculations 
indicate that the corresponding nuclei are characterized by the $T_d^D$
('double-tetrahedral') group of symmetry, exact or approximate.
The corresponding strong shell-gap structure is markedly enhanced by the 
existence of the 4-dimensional irreducible representations of the group in 
question and consequently it can be seen as a {\em geometrical} effect that 
does not depend on a particular realization of the mean-field.
Possibilities of discovering the corresponding symmetry in experiment are 
discussed.

\end{abstract} 

\pacs{PACS numbers: 21.10.-k, 21.60.-n, 21.60.Fw}

\maketitle

%
%

A possibility that atomic nuclei exhibit tetrahedral symmetry - in quantum
physics so far discussed mainly as a property of certain molecules, metal 
clusters or fullerines - has a definite interest for all the related domains 
of physics. While in the above mentioned objects the underlying interactions 
are electromagnetic, the nuclear tetrahedra (pyramid-like nuclei with 'rounded
edges and corners') are expected to be stabilized primarily by the strong
interactions. Within the nuclear mean-field theories, a convenient framework
for discussing this phenomenon is provided by the spontaneous symmetry
breaking mechanism. It is analogous to the one associated with the existence
in nature of  numerous deformed nuclei, e.g. ellipsoidal ones. According to
such a mechanism all nuclei are governed by rotationally-invariant elementary
nucleon-nucleon interactions, yet, for some specific low energy configurations
their total energy becomes lower when the corresponding mean-fields take
non-spherical shapes. Mathematically different but physically analogous
mechanism of spontaneous symmetry breaking is related to a discrete symmetry:
the inversion. The underlying elementary interactions, although
inversion-invariant, do not guarantee that all the resulting low energy
nuclear configurations lead to stable inversion-invariant shapes, and there is
a growing experimental evidence of the existence in nature of the octupole
deformations, usually pear-shape type, cf. e.g. Ref.~\cite{[BNa96]}. It turns
out that the tetrahedral nuclei do break spontaneously both the spherical
symmetry and the symmetry by inversion (see below). 

In the past there has been a number of studies published that address the
question of the non-axially symmetric octupole deformations. Using the 
Strutinsky method and considering a space composed of 2 (quadrupole) + 4
(octupole) + 5 (hexadehapole) + 6 (multipolarity 5) = 17 deformations, the
authors of Ref.~\cite{[XJD94]} have suggested that an ensemble of isomeric 
states of tetrahedral symmetry may exist in the region of light Radium nuclei 
pointing to the importance of the so far neglected $\alpha_{32}$ deformation. 
Using the Hartree-Fock approach in their symmetry-unconstrained variant,
Takami {\em et al.}, Ref.~\cite{[TYM98]}, obtain in some light $Z=N$ nuclei
an  $\alpha_{32}$ instability. In Ref.~\cite{[MYM00]} this and other exotic
octupole deformations were studied in the $^{32}S$ nucleus, while in
\cite{[MYM01]}, a similar hypothesis has been advanced theoretically for a
group of nuclei around $A \sim 70$.

The experimental verification of the discussed phenomenon does not exist so
far. We believe that the mechanism related to $\alpha_{32}$ deformations is
just a 'visible part of an iceberg': a phenomenon whose physical consequences
are much richer than what has been discussed so far. First of all, the
corresponding $T_d^D$ symmetry is nearly unique: only $T_d^D$ and the
octahedral $O_h^D$ point-group symmetries produce in deformed nuclei the
nucleonic level degeneracies higher that 2. More precisely: some states must
carry 2-fold and some 4-fold degeneracies. The corresponding nuclear
hamiltonians are invariant with respect to the very large number of 48
different symmetry elements (in the case of the $O_h^D$ this number would be
96). The depth of the nuclear mean-field potential and the number of its bound
states depend only very weakly on deformation: the 4-fold degeneracy mechanism
at non-zero $\alpha_{32}$ implies larger interspacing and helps in producing
very large shell-gaps that are comparable to or larger than at least some of
the gaps at spherical shapes. Moreover, since the argument is geometrical in
nature the predicted strong shell-gaps propagate all over the periodic table
{\em in a repetitive fashion} independently of a particular realization of the
mean-field approach. This mechanism is far from being an exoticity of a few
nuclei here and there. Its presence is predicted in dozens if not hundreds of
nuclei. Among unique quantum features it should be noted the prediction that
some  nucleonic orbitals should have the expectation value of parity {\em close
to zero} - nearly complete disappearance of the quantum characteristic that is
otherwise dominating in the micro-world of nuclear interactions. Another unique
element foreseen concerns the collective (especially low spin) rotation of the
quantum tetrahedra: the corresponding predicted structures of rotational bands
will be very different for even ($T_d$-group) and odd ($T_d^D$-group) nuclei 
since both the number of the irreps associated with these groups and the
dimensions of the corresponding irreps are very different as well, see e.g.
Refs.~\cite{[Cor84],[Kos63]}.

The mathematical background of our considerations is well known and can be
summarized in a few lines. Consider a nuclear deformed mean-field Hamiltonian
$
      \hat{\mathcal{H}} 
       = 
      \hat{\mathcal{H}}(\vec{r}, \vec{p}, \vec{s};\hat{\alpha})
$
where $\hat{\alpha}$ represents the ensemble of all the deformations 
$\{ \alpha_{\lambda,\mu} \}$
and a group of symmetry $\mathcal{G}$ with the symmetry operators 
$
      \{ {\hat{\mathcal{O}}}_1, 
         {\hat{\mathcal{O}}}_2, \; \ldots 
         {\hat{\mathcal{O}}}_f \}
      \Leftrightarrow
      \mathcal{G}, 
$
so that:
$
      [ \hat{\mathcal{H}}, {\hat{\mathcal{O}}}_k ] =0,
$
for $k=1,2,\; \ldots f$. Suppose that the group in question has irreducible 
representations 
$
      \{ {\mathcal{R}}_1, {\mathcal{R}}_2, \; \ldots {\mathcal{R}}_r \}
$
with the dimensions, respectively:
$
      \{\, d_1, d_2, \; \ldots \; d_r \, \}.
$
Then the eigenvalues $\varepsilon_{\nu}$ of the problem
$
      \hat{\mathcal{H}} \,\Psi_{\nu}
      =
      \varepsilon_{\nu} \,\Psi_{\nu},
$
$\forall \nu$, appear in multiplets: $d_1$-fold degenerate, $d_2$-fold
degenerate, $\ldots$ $d_r$-fold degenerate. Since the nuclear mean-field
Hamiltonians do not depend explicitly on time, it follows that each
eigen-energy must be at least twice degenerate (Kramers theorem). In terms of
irreducible representations of $G$ this property manifests itself for the
deformed nuclei through the presence of  two-dimensional irreducible
representations or pairs of {\em conjugated} one-dimensional ones
\cite{[Kos63]}.

The point-group symmetries of the nuclear mean-field Hamiltonian can be very
often directly connected to the corresponding deformation parameters: if
$\mathcal{R}(\vartheta,\varphi)$ denotes the nuclear surface, expanding it into
a series of spherical harmonics $Y_{\lambda \mu}(\vartheta,\varphi)$ with the
numerical coefficients $\alpha_{\lambda \mu}$ (deformation parameters) provides
a natural classification scheme. Indeed, setting all $\alpha_{\lambda \mu}$ to
zero one obtains a sphere, posing $\alpha_{\lambda=2, \mu=\pm 2, 0} \neq 0$
gives the most important example of the ('ellipsoidal') $D^D_{2h}$-symmetry,
by setting $\alpha_{\lambda=2,3, \mu=0} \neq 0$ we obtain an example of the
octupole-axial ($C^D_{\infty}$) symmetry; many other combinations of the
non-zero deformation parameters may lead to more 'realistic' realizations of
the symmetry groups in nuclei. It can be shown using elementary properties of
the spherical harmonics that in the case of the pure octupole deformations the
following relations between the nuclear shape and the double point-group
symmetries of the fermion Hamiltonians hold:
\begin{itemize}
\item Deformation $\alpha_{30}$: implies the $C_{\infty}^{D}$-symmetry 
      group with infinitely many 1-dimensional irreps., 
      characterized by the so-called K-quantum numbers; 
\item Deformation $\alpha_{31}$: implies the $C_{2v}^{D}$-symmetry
      that generates only 1 two-dimensional irrep;
\item Deformation $\alpha_{32}$: implies the $T_{d}^{D}$-symmetry of 
      2 two-dimensional and 1 \underline{four-dimensional} irreps;
\item Deformation $\alpha_{33}$: implies the $D_{3h}^{D}$-symmetry, 
      and generates 3 two-dimensional irreps.             
\end{itemize}
However, the link between the above scheme and the realistic calculations is not
always direct - especially when the point-groups rich in structure are
concerned. Any isomeric minimum is, within the mean-field theory, associated
with the presence of energy-gaps in the corresponding single particle spectra 
- usually the larger the gap, the more stable the implied equilibrium 
deformation.
\begin{figure}[ht]
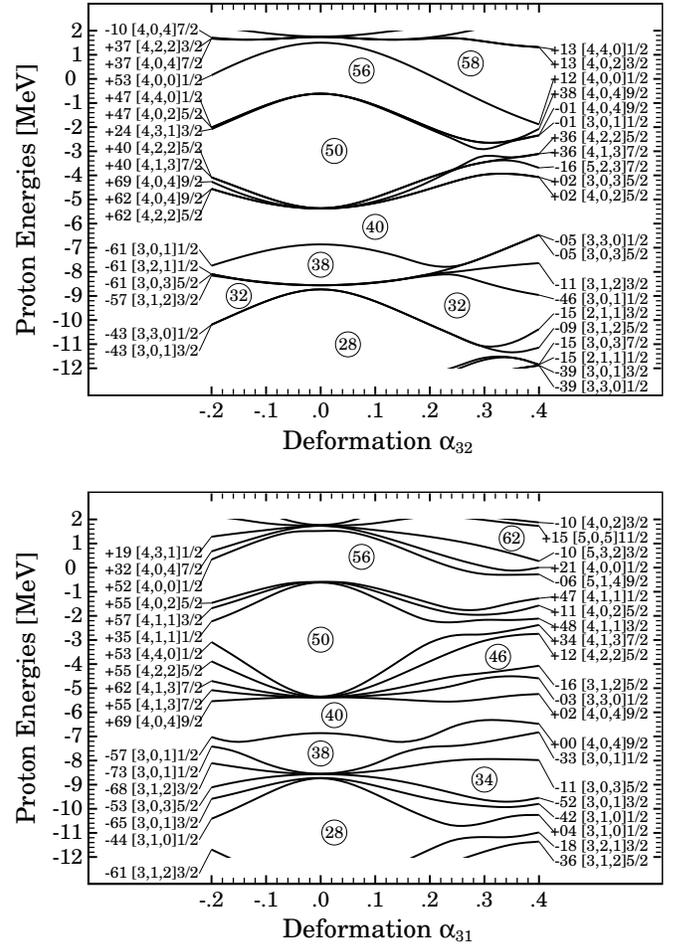
 
  \begin{center} 
    \includegraphics[width=6.0cm,angle=-90]{./90Zr_a32_p_x.epsi}
    \vspace{+0.1truecm}

    \includegraphics[width=6.0cm,angle=-90]{./90Zr_a31_p_x.epsi}
    \caption{\small 
             Results of the realistic calculation of the proton single-particle
             energies in function of the $\alpha_{32}$ deformation corresponding
             to the $T_d^D$ symmetry  (top) compared to
             the analogous dependence in function of $\alpha_{31}$ (bottom,
             $C_{2v}^D$-group). The numbers in front of the 
             Nilsson labels give the expectation values of parity at the 
             extremes of the deformation axes. None of the Nilsson quantum 
             numbers is a good quantum number at tetrahedral deformations: 
             each label gives the full set of quantum numbers of the strongest 
             basis state. 
            (Results obtained using a standard deformed Woods-Saxon 
             potential.)
}                                  
    \label{fig01}                 
  \end{center} 
\end{figure}
Within a given symmetry (given irrep) the non-crossing rule implies that the
single-particle energies tend towards an equidistant  distribution when
deformation increases. If a given group admits only one irreducible
representation each level can carry at most two nucleons. When the
corresponding irreducible representations have higher dimensionality, e.g.
four (the corresponding curves are marked with the double Nilsson labels $[N
n_z \Lambda]\Omega$ in Fig.~\ref{fig01}) more particles may reside on one
single energy level, leading, in some other place of the energy scale to
effectively diminishing the number of levels and possibly increasing the shell
gaps. The features discussed above are followed very closely by the realistic
calculation results of Fig.~\ref{fig01} where, in addition, the extremes on
the horizontal axes have been chosen in such a way that by comparing the
labels on the left- and on the right-hand side of the Figure one can read how
quickly the parity mixing sets in when the deformation increases (the curves
are symmetric with respect to 0). The parity mixing at $\vert \alpha_{32} \vert
\sim 0.15$ is so strong, that the typical calculated parity expectation values
are ($\sim \pm 0.5$).

Several observations deserve emphasizing. First of all, there is a qualitative
difference in the deformation dependence in the two studied cases as predicted
by the considerations based on the point-group symmetries presented above: at
$\vert \alpha_{31} \vert > 0.15$ the level distribution can already be 
considered 'nearly uniform' except for a relatively small gap at $Z=56$ that
{\em decreases} slowly with increasing deformation. In contrast, the spectrum
in function of $\alpha_{32}$ reveals strongly {\em increasing} gaps at $Z=32$,
$\Delta E > 2\;$MeV, at $Z=40$ with $\Delta E \sim 3\;$MeV and a huge  gap at
$Z \to 56, 58$; the latter can be seen as a $\sim 4\;$MeV separation in the
spectrum 'cut across' by a single, usual (i.e. twice degenerate) orbital. These
{\em deformed} gap-sizes are comparable to- or larger than- the strongest 
spherical gaps  at $Z=20, 28, 40$ or even $Z=50$, that are known in the 
medium heavy nuclei; neutron results are similar.

Not all the gaps have an equal impact on the existence (or not) of the well
defined minima on the total energy surfaces. More extended calculations 
whose results will not be presented here in detail can be summarized as follows:
The strongest tetrahedral-symmetry effects appear at proton numbers $Z_t=16$, 
20, 32, 40, $56-58^*$, $70^*$, $90-94^*$, where the asterisks denote the gaps 
that are particularly strong (up to $\sim$3 MeV or so). A clear proton-neutron 
symmetry exists in the calculations leading to the related tetrahedral neutron 
gaps at $N_t=16$, 20, 32, 40, $56-58^*$, $70^*$, $90-94^*$, 112 and 136/142.

\begin{figure}[ht] 
  \begin{center} 
    \leavevmode
    \includegraphics[width=6.0cm,angle=-90]{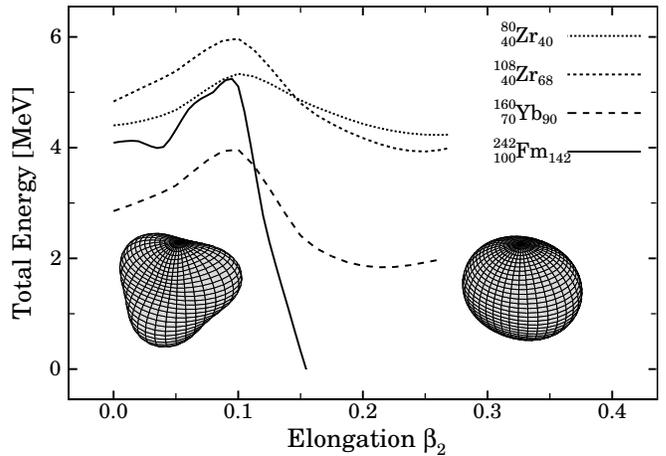}
    \caption{\small 
             Results of the multidimensional minimization of the total nuclear
             energies projected on the quadrupole deformation axis. The gamma
             deformation as well as all other deformations vary along the 
             $\beta_2$-axis following the minimization, for each curve 
             separately. The left-hand side insert shows an exaggerated
             (for better visibility) view of the tetrahedral shape at
             $\alpha_{32}=0.3$, roughly twice the calculated equilibrium
             deformation. The right-hand side insert shows for comparison
             an oblate shape surface at $\beta_2=0.20,\;\gamma=60^0$ i.e. 
             roughly at the calculated equilibria.
}                                  
    \label{fig02}                 
  \end{center} 
\end{figure}

Typically, tetrahedral minima on the total energy surfaces are accompanied by
an oblate- and/or a prolate-symmetry minima. The energy cuts corresponding to
the paths from the tetrahedral minima down to the ground state have been 
calculated in function of increasing $\beta_{2}$ by performing a minimization 
with respect to the $\gamma$ deformation as well as, simultaneously,
$\{\alpha_{3\mu};\;\mu=0,1,2,3 \}$ and
$\{\alpha_{4\mu};\;\mu=0,1,2,3,4 \}$, 10-dimensional minimization, using the 
standard Strutinsky method. These results are presented in Fig.~\ref{fig02}
for  $^{80}_{40}$Zr$_{40}$,  $^{108}_{\;\;40}$Zr$_{68}$,  $^{160}_{\;\;70}$Yb$_{90}$ 
and $^{242}_{100}$Fm$_{142}$ nuclei, whose tetrahedral equilibrium deformations
are calculated at $\alpha_{32}=$0.13, 0.13, 0.15 and 0.11, respectively. 
The right-hand side minima (Fig.~\ref{fig02}) for $^{80}$Zr and $^{160}$Yb
nuclei are at oblate deformations; the corresponding energies visible in the
Figure are, respectively, at 1.4 MeV and $\sim 0.1$ MeV above the  prolate g.s.
minima, the latter not shown in order not to perturb the legibility of the
figure. For the other two nuclei the right-hand side minima  correspond
directly to the prolate ground-states; in the Fm case the tetrahedral minimum
lies paricularly high (7.1 MeV above the g.s.). One can see from the Figure
that the calculated barriers are of the order of 1 MeV, similarly to those 
encountered in the case of the experimentally known prolate/oblate shape 
coexistence. Unfortunately, lack of information about the collective inertia 
parameters makes it impossible to speculate about the isomeric half-lives at 
present.

An experimental identification of proposed $T_d^D$ symmetry may rely on
one or combination of several criteria.

First of all, within the {\em class of the single-particle excitations}, the
presence of the four-fold degeneracies will manifest itself by the presence of
a multitude of particle-hole transitions of close-lying energies. For instance
if both the particle and the hole states are associated to the exact-symmetry
four-fold degenerate levels, one should expect a {\em sixteen-fold} exactly
degenerate multiplet of transitions. (In realistic situations the nuclear
polarization effects are expected to be in general different for various 1p-1h
excitations and the predicted 16-fold degeneracy in the associated decay lines
will only be approximate) If the reference configuration was the tetrahedral
$0^+$ state, the corresponding 16 particle-hole excitations will decay to it.
If as a reference configuration an arbitrary particle-hole excited state built
on the tetrahedral minimum was taken, say of a given spin-parity $I^{\pi}$, a
family of 2p-2h states can be constructed using similar considerations with the
the resulting 16 close-energy transitions feeding this $I^{\pi}$ state. It thus
becomes clear that the non-collective decay spectra associated with the
tetrahedral minima might contain abundantly the approximate 16-plets of
transitions. Although populating and observing such multiplets experimentally
is by far a non-trivial task a good news is that the discussed criterion is a
'yes/no' condition - a well defined effect to seek.
\begin{figure}[ht] 
  \begin{center} 
    \includegraphics[width=6.0cm]{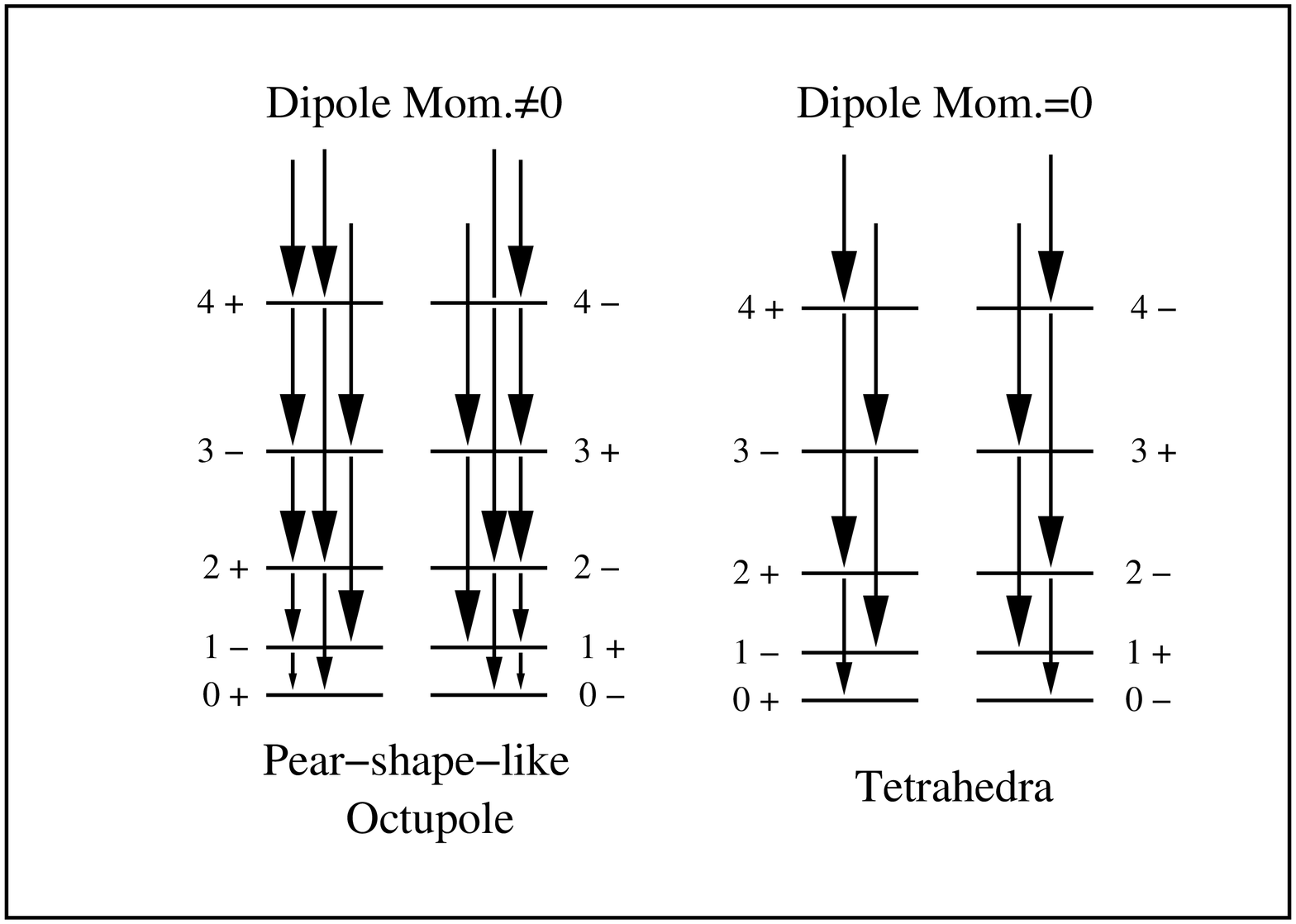}
    \caption{\small 
             Qualitative comparison of the electromagnetic transitions in a
             'pear shape' nucleus, left, and a tetrahedral nucleus,
             right. In the former case the static dipole moments are often
             strong thus implying a presence of the collective inter-band $E1$ 
             transitions in addition to the $E2$ ones. Tetrahedral nuclei 
             generate no static dipole moments 
             and thus the $E1$ transitions should be absent in this case. 
}                                  
    \label{fig03}                 
  \end{center} 
\end{figure}
Within the class of the {\em low-lying collective rotational excitations}
not much is known at present as far as {\em nuclear} tetrahedral quantum rotor 
behavior is concerned. In a formal treatment of the rotational spectra
of the tetrahedral symmetry {\em molecules}, \cite{[Car93]} the 
corresponding rotor Hamiltonians are expanded in terms of tensor operators 
that are constructed out of $\{\hat{I}_x, \hat{I}_y,\hat{I}_z\}$ angular 
momentum operators. The {\em nuclear} rotor Hamiltonians can be constructed
analogously: we have performed the corresponding calculations and the 
preliminary results indicate characteristic degeneracies of the rotor levels
clearly different than those of the 'traditional' ellipsoidal rotors.

Fortunately, qualitative criteria of the 'yes/no' type can be formulated that 
are related to the rotational decay of the tetrahedral nuclei and this in a
nearly model-independent way. The starting point in the considerations is the
so-called {\em simplex-invariance}, i.e. the invariance of the Hamiltonian 
with respect to the product of parity, $\hat{\pi}$ and $\hat{\mathcal{R}}_y$, 
a 180$^o$ rotation about, say, ${\mathcal{O}}_y$-axis: $\hat{\mathcal{S}}$ =
$\hat{\pi} \cdot \hat{\mathcal{R}}_y$. This invariance implies, as discussed in
more detail in Ref.~\cite{[Boh75]}, that the rotational energies form two
parity-doublet sequences (four parity-doublet $E2$-bands) as illustrated
schematically in Fig.~\ref{fig03}. Both the 'usual' axial-octupole shape nuclei
and the tetrahedral nuclei obey to the simplex symmetry and thus must produce
the parity-doublet bands. However, the tetrahedral nuclei, in contrast to the
'usual' octupole-type ones  are {\em not} expected to produce the $E1$
inter-band transitions since the nuclear pyramids, due to their high symmetry,
will not have any significant dipole moments.

A particularly appealing, 'academic case' possibility corresponds to a pure
tetrahedral symmetry with no other multipole deformations present. In such
an ideal case the dipole and quadrupole moments are expected to be zero and the
first non-vanishing moments will have $\lambda=3$. In such a case the
parity-doublet {\em energies} at the right-hand side in Fig.~\ref{fig03}
would be connected by pure $E3$-transitions rather than $E2$-transitions.

In summary: We suggest that the tetrahedral symmetry in nuclei should be
present among many isomeric states throughout the periodic table. We predict 
the proton and neutron numbers for which this effect should be the strongest. 
The four-fold $T_d^D$ degeneracy of related deformed nucleonic orbitals
is a unique feature - only the inversion-conserving octahedral $O_h^D$ symmetry 
may provide in deformed nuclei the degeneracies higher than two (more precisely:
four-fold).
   

\end{document}